\PassOptionsToPackage{table,xcdraw}{xcolor}
\documentclass[sigconf]{acmart}
\usepackage{graphicx}
\usepackage{subcaption}
\AtBeginDocument{%
  \providecommand\BibTeX{{%
    \normalfont B\kern-0.5em{\scshape i\kern-0.25em b}\kern-0.8em\TeX}}}

\setcopyright{acmcopyright}
\copyrightyear{2018}
\acmYear{2018}
\acmDOI{XXXXXXX.XXXXXXX}

\acmConference[Conference acronym 'XX]{Make sure to enter the correct
  conference title from your rights confirmation emai}{June 03--05,
  2018}{Woodstock, NY}
\acmPrice{15.00}
\acmISBN{978-1-4503-XXXX-X/18/06}

\begin{document}

\title{Exploring Musical, Lyrical, and Network Dimensions of Music Sharing Among Depression Individuals}
% Network Music Preferences
\author{Qihan Wang}
\email{qihan.wang@aalto.fi}
\affiliation{%
  \institution{Aalto University}
  \city{Espoo}
  \country{Finland}
  \postcode{02150}
}
\author{Anique Tahir}
\email{artahir@asu.edu}
\affiliation{%
  \institution{Arizona State University}
  \city{Tempe}
  \state{Arizona}
  \country{USA}
  \postcode{85281}
}

\author{Zeyad Alghamdi}
\email{zalgham1@asu.edu}
\affiliation{%
  \institution{Arizona State University}
  \city{Tempe}
  \state{Arizona}
  \country{USA}
  \postcode{85281}
}
  
\author{Huan Liu}
\email{huanliu@asu.edu}
\affiliation{%
  \institution{Arizona State University}
  \city{Tempe}
  \state{Arizona}
  \country{USA}
  \postcode{85281}
}

\renewcommand{\shortauthors}{Trovato and Tobin, et al.}

\begin{abstract}
Depression has emerged as a significant mental health concern due to a variety of factors, reflecting broader societal and individual challenges. Within the digital era, social media has become an important platform for individuals navigating through depression, enabling them to express their emotional and mental states through various mediums, notably music. Specifically, their music preferences, manifested through sharing practices, inadvertently offer a glimpse into their psychological and emotional landscapes. This work seeks to study the differences in music preferences between individuals diagnosed with depression and non-diagnosed individuals, exploring numerous facets of music, including musical features, lyrics, and musical networks. The music preferences of individuals with depression through music sharing on social media, reveal notable differences in musical features and topics and language use of lyrics compared to non-depressed individuals. We find the network information enhances understanding of the link between music listening patterns. The result highlights a potential echo-chamber effect, where depression individual's musical choices may inadvertently perpetuate depressive moods and emotions. In sum, this study underscores the significance of examining music's various aspects to grasp its relationship with mental health, offering insights for personalized music interventions and recommendation algorithms that could benefit individuals with depression.

\end{abstract}




\received{20 February 2007}
\received[revised]{12 March 2009}
\received[accepted]{5 June 2009}

\maketitle

\section{Introduction}
Depression is a prevalent mental disorder characterized by persistent feelings of sadness, a diminished interest, or a lack of pleasure in daily activities \cite{gotlib2008handbook}. Depression impacts approximately 3.8\% of the global population across all ages, genders, and cultural groups \cite{leubner2017reviewing}, posing a significant worldwide challenge \cite{WHO}. Distinct from regular mood fluctuations, depression is characterized by its prolonged duration, and may manifest in feelings of excessive guilt or low self-worth, loss of interest or pleasure, hopelessness about the future, disturbed sleep or appetite, and poor concentration \cite{woody2017systematic, evans2018socio}. With the development of social media, individuals experiencing depression tend to express their thoughts and share their feelings with their peers and audiences online \cite{andalibi2017sensitive}. The shared content includes a variety of forms, such as text, photos, news and music. 

Music, intertwining with daily lives, enables individuals to regulate emotions, maintain interpersonal relationships, and express their identities \cite{lee2011understanding, johnson2018click}. As people connect their own memories and thoughts with music \cite{krumhansl2002music}, music sharing can serve as a mirror to reflect people's moods and feelings towards life events \cite{robinson2005deeper, greasley2006music}. Building on this, individuals with depression may perceive music differently and exhibit distinct musical preferences. Previous psychology research has shown that individuals with a tendency towards depression demonstrated a preference for sad music \cite{garrido2015music}, and individuals who are diagnosed by depression perceive music as conveying more negatively emotion \cite{eerola2011comparison}. However, most of the previous research on music and depression focused on the emotion of the songs, ignoring other musical features that can be quantified or measured by rhythmic and structural characteristics, such as tempo, energy, mode and acousticness. In addition, they often overlook elements beyond musical features, such as lyrics and associated networks.
% in this work we look at 2 groups one is assigned depression based om xyz, and we have the non idagonsed or control group those that are not self daignosed

In sum, while prior investigations have primarily studied the emotional aspects of songs, the broader spectrum of musical attributes, lyrics, and associated networks largely remains unexplored. This significant gap presents a chance to explore in more detail the differences of music preferences between individuals with and without depression. Given this context, our study takes a broad approach, carefully examining not just various musical features, but also delving into lyrics and network characteristics among social media users. Based on a publicly available dataset \cite{Kaggle} that encompasses music sharing data from users who have self-expressed that they have been diagnosed with depression, together with data from randomly-selected control users (non-depression users), we studied 4 research questions:

%Such analysis is pivotal for enhancing our understanding of the online behavior of users experiencing depression and provides valuable insights for the development of music recommendation algorithms that are attuned to the needs of individuals navigating mental health disorders.

\textbf{RQ1: Do musical features of music sharing differ between depression and non-depression individuals?} By collecting musical features of each song from Spotify API \cite{SpotifyAPI}, we discern any notable variations between the depression and non-depression cohorts. We found that songs shared by depression are significant higher in energy, mode, acousticness, instrumentalness, and duration, and lower in speechiness, time signature, popularity and danceability. 

\textbf{RQ2: What are specific artists or genres that are commonly shared by depression and non-depression individuals?} We presented qualitative visualizations for artists and genres shared by depression groups and control groups, and leveraged TF-IDF to present  artists and genres that are uniquely shared by depression groups.

\textbf{RQ3: Do topics and language use in lyrics of music sharing differ between depression and non-depression individuals?} We conducted a multi-faceted lyric level analysis. Our result showed that music shared by individuals with depression have higher proportion in the topic of emotional reflection and religious language, yet lower proportion in the topic of explicit (RAP) languages and personal relationships. Furthermore, individuals with depression prefer lyrics with shorter sentences.

\textbf{RQ4: What are the network relationships regarding music shared by depression and non-depression individuals?} Based on Spotify Playlist Dataset \cite{SpotifyPlaylist}, we categorized playlists as depression/control playlists by the amount of songs that shared by depression users and conducted community detection for all playlists. Our result indicated that the depression playlists cluster together, suggesting a network connection between depression songs. Combined and coherent with the previous lyric analysis, we found that emotional reflection and religious or biblical language are dominant topics in depression playlist communities.

In synthesizing our findings, we discern crucial implications across two pivotal dimensions. Firstly, our results demonstrated social and cognitive characteristic features of individuals with depression in relation to their music preferences. Secondly, we delved into the consequences engendered by these specific musical preferences. We found that the music preference of individuals with depression potentially encapsulates users within a depression Echo-Chamber Effect, where the music exacerbates prevailing depressive moods and emotions. This explorative analysis not only heightened our understanding of the nuanced relationships between music preferences and mental health but also illuminated pathways for the development of music recommendation algorithms that were attuned to the needs of individuals navigating mental health disorders.

\section{Related Work}
Music, an integral component of human interaction, has seen its influence magnified due to advances in digital technology, especially on platforms like social media \cite{kassabian2013ubiquitous, jones2000music}. These platforms have become central stages where music is shared, offering insights into individuals' identity and self-presentation \cite{johnson2018click, valverde2022online}. In this work, we focus on its relationship with mental states, particularly depression. Specifically, we explore the multifaceted connections between music sharing and depression, including musical features, lyrics, and social network relationships. This section aims to provide an overview, drawing from existing research to shape our discourse on these interconnected themes.

%\subsection{Music Sharing on Social Media}

%Digital techniques highly increased the ubiquity and the social role of music \cite{kassabian2013ubiquitous, jones2000music}. Music has maintained its central role for daily communicative practices on social media \cite{valverde2022online}. The main motivation of sharing music on social media is self-presentation for imagined audience, including presenting the actual-self and the ideal-self \cite{johnson2018click, valverde2022online}. 

\subsection{Music Features and Its Multifaceted Connection to Individuals}

Our study aims to incorporate a diverse array of musical features, providing a more comprehensive understanding of the music preferences of individuals with depression. Spanning from rhythmic and structural to timbral characteristics, these features serve as quantitative representations of various dimensions of music, enabling a wide range of applications such as music recommendation, playlist generation, and exploratory musicological research \cite{SpotifyAPI}. These features encompass various attributes such as danceability, duration, energy, instrumentalness, liveness, loudness, speechiness, tempo, and valence \cite{barone2017acoustic,greenberg2016song}. Each feature encapsulates a distinct aspect of a song, elucidating its various musical and acoustical properties; for instance, danceability refers to how suitable a track is for dancing, often through examining elements like beat regularity and strength, and valence serves to describe the musical positiveness conveyed by a track, distinguishing between sonically positive (e.g., happy, cheerful) and negative (e.g., sad, depressive) emotional expressions. Moreover, research have found that there's a notable correlation between these musical features and individuals' traits: individual's personality aligns with their affinity or preference for specific musical features, including mode, tempo, and dynamics  \cite{flannery2021musical,ferwerda2015personality}.

\subsection{Overview of Depression and Music}

Music preference is intricately linked with an individual's conditions and traits, including mental health conditions \cite{lester1996music}, personality \cite{delsing2008adolescents}, and intelligence \cite{bonetti2016intelligence}. Specifically, when examining the connection with depression, past research has predominantly centered on musical genres and sentiments. In the context of genres, prior studies suggest that listening to soul music (hip hop and R\&B) is a predictor of lower depression levels \cite{miranda2008personality}. As for sentiment, previous research has demonstrated that individuals diagnosed with depression perceive music as more negatively valence \cite{eerola2011comparison}. Such individuals exhibit a tendency to gravitate towards negatively valenced music, mirroring their chronic mood state \cite{wilhelm2013blue}.  Moreover, listening to sad music can significantly increase the depressed feelings of individuals with depression \cite{garrido2015music}.

Nevertheless, music has also been identified as a therapeutic tool for individuals with depression. Studies indicate that music listening can lead to a notable reduction in depression levels \cite{maratos2008music}. This therapeutic approach, termed music therapy, has been corroborated as an efficacious treatment for depression \cite{erkkila2011individual}.

\subsection{Overview of Depression and Social Media}

Recent research increasingly harnesses user-generated content on social media to explore mental health-related issues, generally employing two primary approaches: (1) analyzing the behavior patterns, cognition, and emotional aspects of mental health disorders \cite{xu2016understanding}, and (2) developing automatic detection systems for mental health problems or anomalies \cite{de2013predicting, guntuku2017detecting}. These studies broadly utilizes existing tools and methods from data mining, natural language processing, social network analysis, and machine learning. Within this framework, a spectrum of mental health disorders, including major depressive disorders \cite{de2013predicting}, eating disorders \cite{frieiro2021social}, and ADHD \cite{guntuku2019language}, have been explored. Our work is anchored in the first approach, and it aims to discern the variance in music-sharing behaviors between individuals diagnosed with depression and those who aren't.

Building upon the foundation of the first approach, earlier research focusing on depression and its manifestation on social media has delved into aspects like social interaction, linguistic expression, and emotional displays of affected users. From a sociological standpoint, it's been observed that individuals with depression tend to manifest a decrease in social activity and form highly clustered ego-networks \cite{de2013social}. From linguistical perspectives, it has been shown that individuals with depression express more distorted thinking, and their online language is generally deemed depressogenic \cite{bathina2021individuals}. Notably, the linguistic patterns and topic-centric vocabulary of depressed individuals significantly deviate from their non-depressed counterparts \cite{rissola2022mental}.  On the emotional front, it's been found that self-reported depression is related to the usage of personal and negative pronoun language categories \cite{liu2022relationship}. Furthermore, individuals with depression form more negative sentences from scrambled words. In other words, they tend to form more pessimistic sentences than non-depressed participants, even when they are not in a negative mood \cite{wenzlaff1998unmasking}. Moreover, from topic perspectives, people with depression have amplified relational and medicinal concerns and express greater religious involvement \cite{de2013social}. While previous research has established a discrepancy in the behavior and expression between individuals with and without depression on social media, spanning sociality, language expression, and emotion, a gap persists in understanding the domain of music preferences. Our study aspires to bridge this gap, aiming to discern how music sharing practices diverge between these two groups, consequently presenting a fresh lens to the existing discourse.

%In sum, the behavior and expression on social media are highly different between people with and without depression, including sociality, language expression and emotion. To extend works on the analysis of differences, our study contributes to analyzing the music sharing differences between people with and without depression.

%Moreover, there are significant difference musical feature preference between people with different IQ \cite{chamorro2007personality}.

% Music sharing on social media
% Overview of Depression: music and social media
% Music features and Its Multifaceted Connection to Individuals

\section{Evaluation \& Results}

In this section, we present the findings obtained from our analysis focusing on three primary domains: musical variables, lyrics and network. Our objectives were to: (1) identify significant differences in musical variables, artists and genres between depression and non-depression groups by utilizing statistical analysis (2) present artist and genres differences qualitatively by word cloud (3) understand the difference of topics and language use of the lyrics between depression and non-depression groups by LDA and LIWC (4) explore the connection of depression playlists by leveraging community detection.

\subsection{Data Collection}

To achieve our research objectives, we carefully selected a dataset that aligns with the intersection of mental health and musical engagement, which is a crucial part of our investigation. But also aiding us with a notable advantage, that is, the inclusion of a depression and a control group, which provides a foundational comparison and paves the way for a deeper and more comprehensive analysis of the interplay between music-sharing behaviors and mental states in a digital context. Therefore, we sourced an anonymized publicly available dataset that encompasses social media information ~\cite{Kaggle}. In our selected dataset, the 'Depression Group' is characterized by users who have self-identified with a depression diagnosis on Twitter, while the 'Control Group', also known as 'Non-Depression Group', includes those who never tweeted depression or any other mental disorders. Every entry in this dataset provides specifics like the song's title, artist, and cleaned-up lyrics. There's also a marker to identify whether the song-sharer publicly disclosed a depression diagnosis on Twitter. We further ensured a balanced representation of music-sharing posts between both groups. For a richer analysis of the link between music-sharing and depression, we tapped into the Spotify API~\cite{SpotifyAPI,panda2021does} to extract musical attributes for each shared piece. This extraction process yielded 14 numerical variables, which are described in table \ref{tab:numerical_variables}. Each variable reveals a unique facet of the musical piece. Given the wide spectrum of genres a song can belong to, the collective attributes present a panoramic perspective on musical styles and their potential implications for understanding an individual's psychological state.

\begin{table}[h]
\centering
\small
\caption{Description of Numerical Variables from Spotify API}
\begin{tabular}{|l|p{5cm}|}
\hline
\textbf{Variable} & \textbf{Description} \\
\hline
Danceability & Danceability is a measure from 0.0 to 1.0 which describes how suitable a track is for dancing based on a combination of musical elements including tempo, rhythm stability, beat strength, and overall regularity. \\
\hline
Energy & Energy is a measure from 0.0 to 1.0 and represents a perceptual measure of intensity and activity. Typically, energetic tracks feel fast, loud, and noisy. \\
\hline
Key & The key the track is in, using standard Pitch Class notation. \\
\hline
Loudness & Overall loudness of a song in decibels (dB). \\
\hline
Mode & Modality of a track; 1 represents major and 0 represents minor. \\
\hline
Speechiness & Speechiness detects the presence of spoken words in a track. The more exclusively speech-like the recording (e.g. talk show, audio book, poetry), the closer to 1.0 the attribute value.  \\
\hline
Acousticness & A confidence measure from 0.0 to 1.0 of whether the track is acoustic. \\
\hline
Instrumentalness & Predicts whether a track contains no vocals. \\
\hline
Liveness & Detects the presence of an audience in the recording. Higher liveness values represent an increased probability that the track was performed live. \\
\hline
Valence & A measure from 0.0 to 1.0 describing the musical positiveness conveyed by a track. \\
\hline
Tempo & The overall estimated tempo of a track in beats per minute (BPM). In musical terminology, tempo is the speed or pace of a given piece and derives directly from the average beat duration. \\
\hline
Duration\_ms & Duration of the song in milliseconds. \\
\hline
Time\_Signature & Estimated overall time signature of a track, which specify how many beats are in each bar \\
\hline
Popularity & Measure of the track's popularity based on number of plays. \\
\hline
\end{tabular}
\label{tab:numerical_variables}
\end{table}

For network analysis, we leverage the Spotify Million Playlist Dataset~\cite{SpotifyPlaylist}, which contains titles for songs and playlists created by users on the Spotify platform between January 2010 and October 2017. We aimed to examine if playlists with songs linked to depression cluster distinctively compared to those without such songs, aiming to shed light on the intricate relationship between music preferences and mental health.

\subsection{Musical Features}
% The graph is in the "Descriptive Statistics for Musical Features" files
% descriptive analysis table for musical features
In addressing \textbf{RQ1} through our exploration of musical features, we aimed to identify any significant differences in musical variables between depression and non-depression groups. Our null hypothesis, denoted as \(H_0\), posits no significant difference in the evaluated musical attributes between the two cohorts. We subjected this hypothesis to empirical verification employing a suite of statistical tests. We found that there are significant differences between depression and control group in danceability ($MD=-0.041$, $p<0.001$), energy ($MD=-0.012$, $p<0.001$), speechiness ($MD=-0.030$, $p<0.001$), acousticness ($MD=0.009$, $p<0.05$), instrumentalness ($MD=0.022$, $p<0.001$), Time Signature ($MD=-0.026$, $p<0.001$), popularity ($MD=-1.000$, $p<0.001$). Moreover, the duration of the song is significantly longer in depression group ($MD=8875.318$, $p<0.001$). On the other hand, there are significantly more major mode and less minor mode in depression group ($MD=0.044$, $p<0.001$).Additionally, there is no difference in key, loudness, liveness, valence, and tempo.

\begin{table}[]
\caption{Table 2: Statistical Analysis of Musical Difference}
\begin{tabular}{
>{\columncolor[HTML]{FFFFFF}}l 
>{\columncolor[HTML]{FFFFFF}}l 
>{\columncolor[HTML]{FFFFFF}}l 
>{\columncolor[HTML]{FFFFFF}}l 
>{\columncolor[HTML]{FFFFFF}}l }
\hline
{\color[HTML]{000000} Variable}         & {\color[HTML]{000000} T-value}    & {\color[HTML]{000000} df}    & {\color[HTML]{000000} MD}         & {\color[HTML]{000000} p-value} \\ \hline
{\color[HTML]{000000} danceability}     & {\color[HTML]{000000} -17.070585} & {\color[HTML]{000000} 18262} & {\color[HTML]{000000} -0.041}  & {\color[HTML]{000000} \textbf{0.000$^{***}$}}              \\
{\color[HTML]{000000} energy}           & {\color[HTML]{000000} 4.03491}    & {\color[HTML]{000000} 18262} & {\color[HTML]{000000} 0.012}   & {\color[HTML]{000000} \textbf{0.000$^{***}$}}       \\
{\color[HTML]{000000} key}              & {\color[HTML]{000000} -0.127663}  & {\color[HTML]{000000} 18262} & {\color[HTML]{000000} -0.007}  & {\color[HTML]{000000} 0.898}       \\
{\color[HTML]{000000} loudness}         & {\color[HTML]{000000} -2.202757}  & {\color[HTML]{000000} 18262} & {\color[HTML]{000000} -0.102}  & {\color[HTML]{000000} \textbf{0.028$^*$}}       \\
{\color[HTML]{000000} mode}             & {\color[HTML]{000000} 6.087954}   & {\color[HTML]{000000} 18262} & {\color[HTML]{000000} 0.044}   & {\color[HTML]{000000} \textbf{0.000$^{***}$}}              \\
{\color[HTML]{000000} speechiness}      & {\color[HTML]{000000} -18.630115} & {\color[HTML]{000000} 18262} & {\color[HTML]{000000} -0.030}  & {\color[HTML]{000000} \textbf{0.000$^{***}$}}              \\
{\color[HTML]{000000} acousticness}     & {\color[HTML]{000000} 2.229929}   & {\color[HTML]{000000} 18262} & {\color[HTML]{000000} 0.009}   & {\color[HTML]{000000} \textbf{0.026$^*$}}       \\
{\color[HTML]{000000} instrumentalness} & {\color[HTML]{000000} 8.837124}   & {\color[HTML]{000000} 18262} & {\color[HTML]{000000} 0.022}   & {\color[HTML]{000000} \textbf{0.000$^{***}$}}              \\
{\color[HTML]{000000} liveness}         & {\color[HTML]{000000} 0.815591}   & {\color[HTML]{000000} 18262} & {\color[HTML]{000000} 0.002}   & {\color[HTML]{000000} 0.415}       \\
{\color[HTML]{000000} valence}          & {\color[HTML]{000000} -0.700019}  & {\color[HTML]{000000} 18262} & {\color[HTML]{000000} -0.002}  & {\color[HTML]{000000} 0.484}       \\
{\color[HTML]{000000} tempo}            & {\color[HTML]{000000} 1.768637}   & {\color[HTML]{000000} 18262} & {\color[HTML]{000000} 0.778}   & {\color[HTML]{000000} 0.077}       \\
{\color[HTML]{000000} duration\_ms}     & {\color[HTML]{000000} 7.613851}   & {\color[HTML]{000000} 18262} & {\color[HTML]{000000} 8875.318} & {\color[HTML]{000000} \textbf{0.000$^{***}$}}              \\
{\color[HTML]{000000} time\_signature}  & {\color[HTML]{000000} -4.905537}  & {\color[HTML]{000000} 18262} & {\color[HTML]{000000} -0.026}  & {\color[HTML]{000000} \textbf{0.000$^{***}$}}       \\
{\color[HTML]{000000} popularity}       & {\color[HTML]{000000} -3.622495}  & {\color[HTML]{000000} 18262} & {\color[HTML]{000000} -1.000}  & {\color[HTML]{000000} \textbf{0.000$^{***}$}}       \\ \hline
\end{tabular}
%\begin{itemize}
\text{$^*$ refers to the significance(}$^*$ $p < 0.05$ , $^{**}$ $p<0.01$ , $^{***}$ $p < 0.001$)
%\end{itemize}
\end{table}
\raggedbottom

%i have comments on that part above
%Word Cloud
In addition to the quantitative analysis for artists and genres for answering \textbf{RQ2}, we also employ qualitative visualizations to intuitively represent the prevalent elements in our dataset. 

\begin{figure*}[htbp]
    \centering
    
    \begin{subfigure}[b]{0.3\textwidth}
        \includegraphics[width=\textwidth]{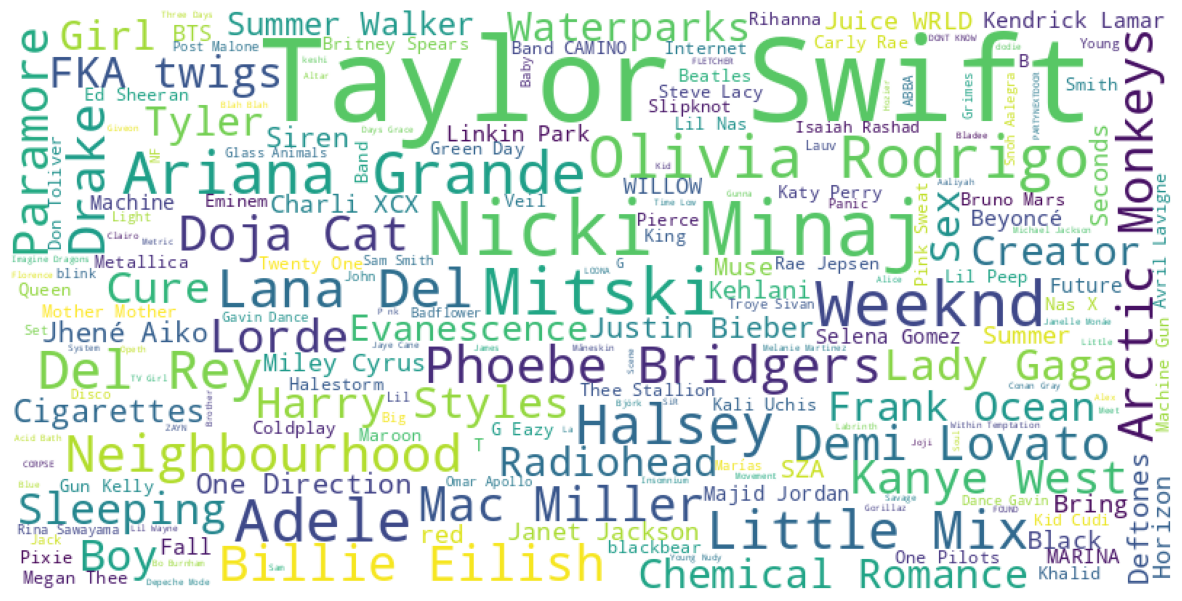}
        \caption{Artist of Depression Group}
        \label{fig:depression_artist}
    \end{subfigure}
    \hfill % add some horizontal spacing between the two images
    \begin{subfigure}[b]{0.3\textwidth}
        \includegraphics[width=\textwidth]{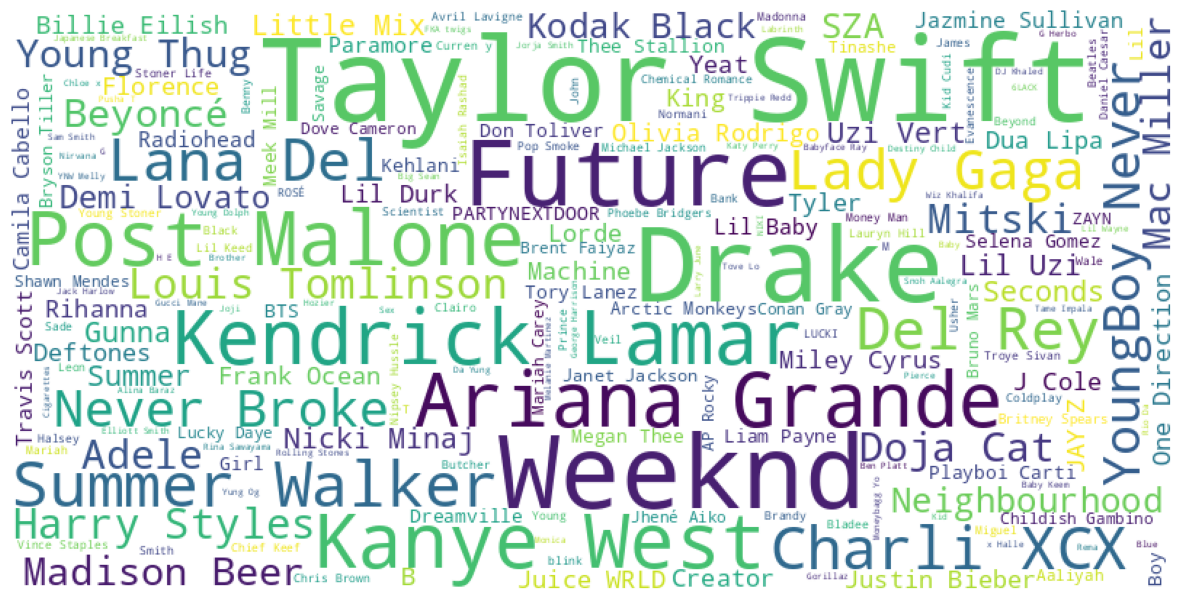}
        \caption{Artist of Control Group}
        \label{fig:control_artist}
    \end{subfigure}
    \hfill
    \begin{subfigure}[b]{0.3\textwidth}
        \includegraphics[width=\textwidth]{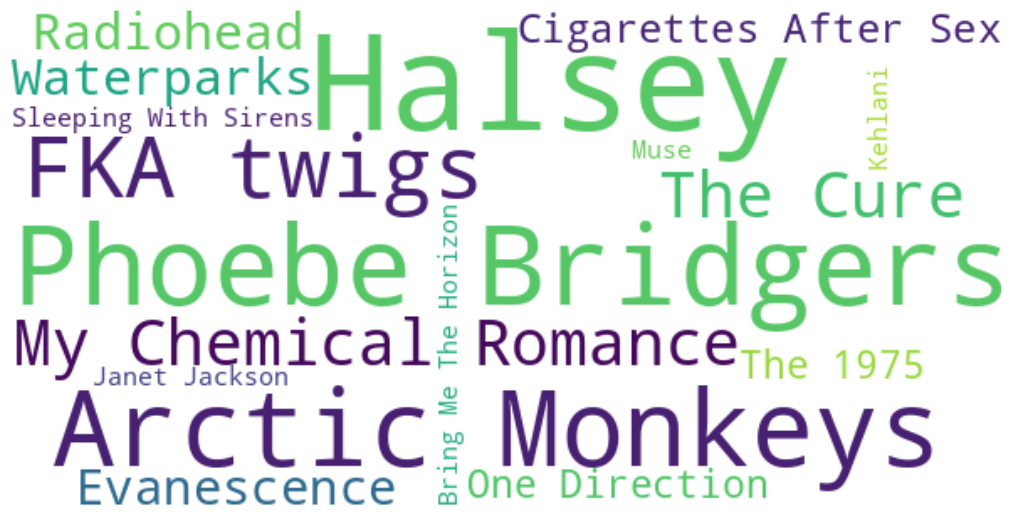}
        \caption{Unique Artist Items in Depression}
        \label{fig:unique_artist}
    \end{subfigure}
    
    \vspace{1em} % add some vertical spacing between the two rows
    
    \begin{subfigure}[b]{0.3\textwidth}
        \includegraphics[width=\textwidth]{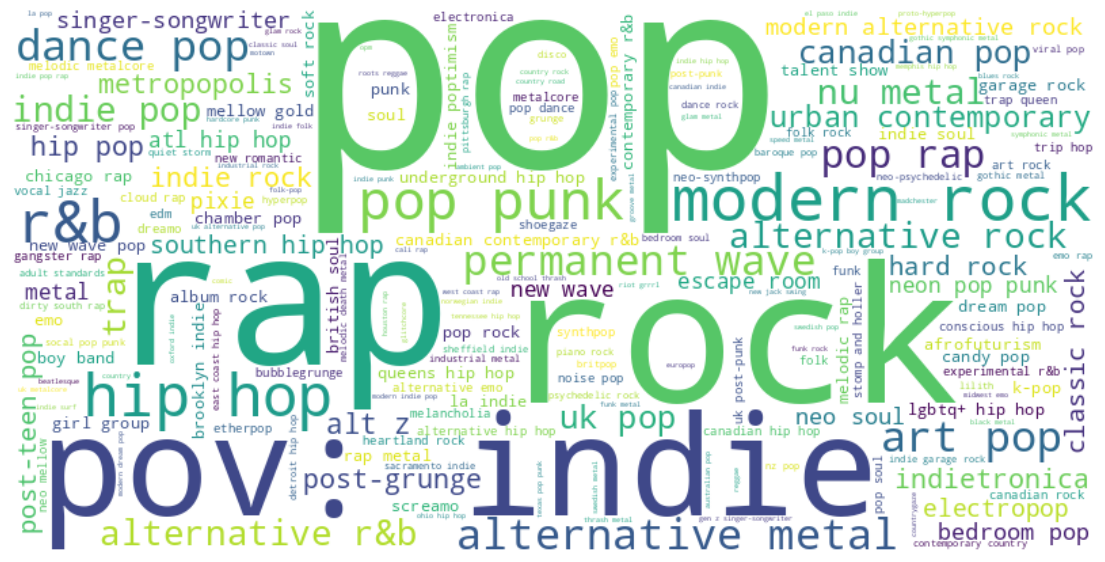}
        \caption{Genre of Depression Group}
        \label{fig:depression_genre}
    \end{subfigure}
    \hfill
    \begin{subfigure}[b]{0.3\textwidth}
        \includegraphics[width=\textwidth]{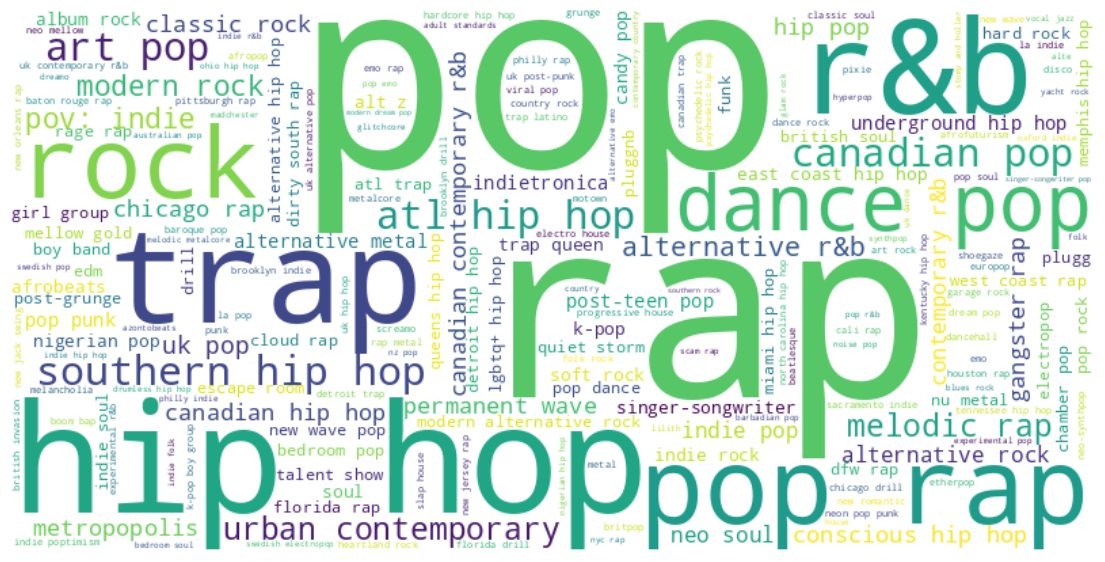}
        \caption{Genre of Control Group}
        \label{fig:control_genre}
    \end{subfigure}
    \hfill
    \begin{subfigure}[b]{0.3\textwidth}
        \includegraphics[width=\textwidth]{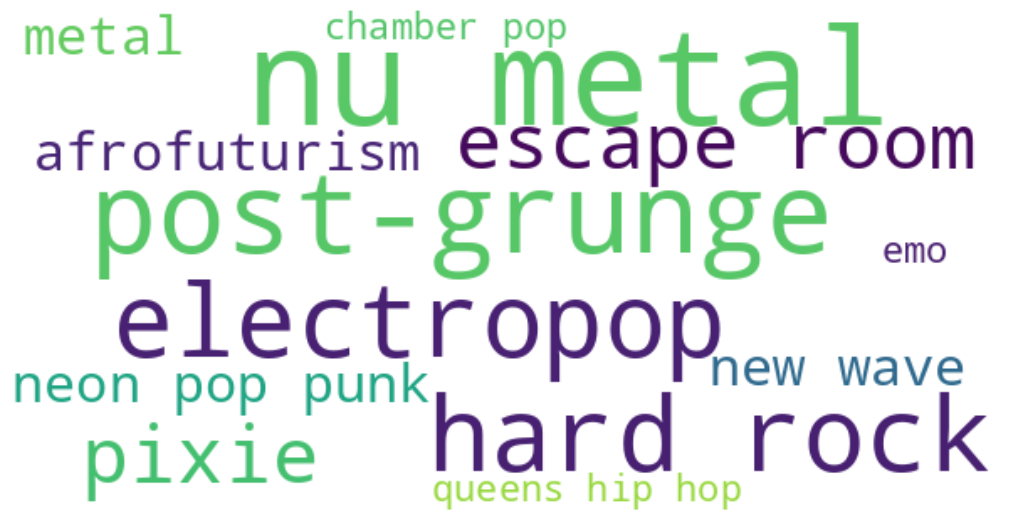}
        \caption{Unique Genres Items in Depression}
        \label{fig:unique_genres}
    \end{subfigure}

      %\vspace{1em} % add some vertical spacing between the two rows
    
    %\begin{subfigure}[b]{0.45\textwidth}
       % \includegraphics[width=\textwidth]{Sections/Word Cloud/Depression_Lyrics.png}
      %  \caption{Lyrics of Depression Group}
      %  \label{fig:depression_lyrics}
   % \end{subfigure}
   % \hfill
   % \begin{subfigure}[b]{0.45\textwidth}
       % \includegraphics[width=\textwidth]{Sections/Word Cloud/Control_Lyrics.png}
      %  \caption{Lyrics of Control Group}
      %  \label{fig:control_lyrics}
   % \end{subfigure}
    
    \caption{Qualitative Visualization of Artist and Genres}
    \label{fig:wordcloud}
\end{figure*}

Specifically, we construct word clouds of depression and control group for artists (figure \ref{fig:depression_artist} and \ref{fig:control_artist}) and genres (figure \ref{fig:depression_genre} and \ref{fig:control_genre}). By leveraging TF-IDF, we present the unique artist \ref{fig:unique_artist} and genres \ref{fig:unique_genres} in depression group. 

%graphs
% TODO subfigure 
\subsection{Lyrics Analysis}
Although the analysis of musical features have provided invaluable insights, there remains an intricate layer of understanding that can only be unraveled through an examination of lyrics. Lyrics, imbued with semantic and emotional content, serve as a conduit for artists to express intricate sentiments, and for listeners to resonate with these expressions. The lyrical content can amplify, complement, or even contrast the emotional tenor conveyed by musical elements. Given that individuals with depression may be particularly attuned to the emotional and thematic content of lyrics(add citation), a granular analysis at the lyric level is paramount. It facilitates a more comprehensive understanding of the specific themes, narratives, and emotional expressions that resonate with this demographic. This, in turn, could unveil nuanced patterns and correlations, offering a richer, multidimensional perspective on the intricate interplay between music, lyrics, and mental health, and potentially informing targeted, lyrics-based interventions for individuals with depression.

In pursuit of \textbf{RQ3} that focusing on the lyrics differences, we conducted multi-faceted lyric level analysis. We use traditional topic modelling to distinguish clusters at a linguistic level. In addition, we use LIWC based approach to analyze the lyrics at a more emotional, cognitive, and structural level.

\subsubsection{Topic Modeling}

In our lyrics analysis, we employ Latent Dirichlet Allocation (LDA) to identify latent topics from the corpus of song lyrics, aiming to unveil thematic distinctions between the depression and control groups. LDA is a generative probabilistic model in which each document (in this case, a set of lyrics) is represented as a random mixture of latent topics \cite{blei2003latent}. Each topic is characterized by a distribution over words in the vocabulary.

Given a corpus of lyrics $D$, each song's lyrics $d \in D$ is modeled as a finite mixture over an underlying set of topics. Each topic $k$, represented as $\beta_k$, is a distribution over the vocabulary of words. The topics are assumed to have been drawn from a Dirichlet distribution with parameter $\eta$:

\[
\beta_k \sim \text{Dir}(\eta), \quad k = 1, \ldots, K.
\]

Each document (or set of lyrics) $d$ is associated with a vector of topic proportions $\theta_d$, which is drawn from a Dirichlet distribution with parameter $\alpha$:

\[
\theta_d \sim \text{Dir}(\alpha), \quad d = 1, \ldots, D.
\]

Each word $w_{di}$ in document $d$ is generated by first sampling a topic $z_{di}$ from the multinomial distribution parameterized by $\theta_d$, and then sampling the word from the multinomial distribution parameterized by $\beta_{z_{di}}$:

\[
z_{di} \sim \text{Multinomial}(\theta_d), \quad w_{di} \sim \text{Multinomial}(\beta_{z_{di}}).
\]

We use the estimated topic-word distributions and document-topic distributions to analyze the thematic content of lyrics in both the depression and control groups. Statistical tests, such as the Proportion Z-Test, are then applied to evaluate the significance of the observed differences in topic proportions between the two groups. The most common 15 topics in the lyrics are shown in table \ref{tab:LDA}

The Proportion Z-Test is employed to assess whether the proportions of the topic differ significantly between depression and control group. Compared to the control group, there is a statistically significant higher proportion of Music Recording($p<0.05$, $MD=0.002$), Religious or Biblical Language ($p<0.01$, $MD=0.005$), Everyday Life ($p<0.01$, $MD=0.0039$), Entertainment Industry($p<0.01$, $MD=0.002$), Recollection and Creativity($p<0.05$, $MD=0.003$), Everyday Experiences ($p<0.001$, $MD=0.036$), Emotional Reflection ($p<0.001$, $MD=0.093$), Everyday Conversations($p<0.01$, $MD=0.004$) in the depression group, and there is a statistically significant lower proportion of Explicit or RAP Language ($p<0.001$, $MD=-0.084$), Expressive Language($p<0.001$, $MD=-0.0403$), Musical Artists and Collaborations($p<0.001$, $MD=-0.007$), and Personal Relationships ($p=0.01$, $MD=-0.010$) in the depression group.

\begin{table*}[t]
\caption{Topic Modeling Result}
\begin{tabular}{lll}
\hline
Topic & Interpretaion                                                     & Most representative topic                \\
\hline
0     & Music Recording                                                   & studio, recorded, nyc, gun, machine      \\
1     & Religious or Biblical Language                                    & god, king, thou, said, man               \\
2     & Everyday Life                             & people, going, year, like, school, work  \\
3     & Entertainment Industry                     & taylor, award, music, batman, man        \\
4     & Love and Relationships                     & baby, love, like, want, girl             \\
5     & Political or Philosophical Discussion      & great, man, people, power, state         \\
6     &  Conversational Language                   & say, tell, know, dont, like              \\
7     &  Explicit or RAP Language                  & shit, bitch, fuck, nigga, money          \\
8     &  Recollection and Creativity                & remember, love, night, christmas, summer \\
9     &  Expressive Language          & dont, know, love, like, want             \\
10    &  Individual Names and Personal Interactions & stephen, zoe, man, old, say              \\
11    &  Time and Personal Relations                & love, january, april, young, july        \\
12    &  Timeframes and Specific Events            & january, march, april, february, june    \\
13    &  Legal and Constitutional Discourse         & state, court, law, power, congress       \\
14    &  Everyday Experiences                       & like, come, life, home, hand             \\
15    &  Musical Artists and Collaborations        & feat, lil, west, kanye, travis           \\
16    &  Emotional Reflection                       & time, away, heart, love, way             \\
17    &  Explicit or Medical Words                 & nigga, health, secretary, bitch, service \\
18    &  Personal Relationships                     & love, like, young, boy, girl             \\
19    &  Everyday Conversations                     & like, said, time, day, know             \\
\hline
\end{tabular}
\label{tab:LDA}
\end{table*}

\subsubsection{LIWC Across Single-Occurrence and Repeated Songs}

%In the preprocessing phase, the original dataset comprising 35,065 records and 20,882 unique IDs was filtered to retain only those entries with unique IDs, resulting in a refined dataset of 15,518 records.

To explore the characteristics of songs, we employed the Linguistic Inquiry and Word Count (LIWC) tool \cite{boyd2022liwc}. This tool provides features that capture the text's linguistic, emotional, cognitive, and structural elements. We designed two subsets of the original dataset, one for each of the depression and control groups, while eliminating overlapping songs to maintain methodological rigor.

For a given lyric \( L \), LIWC generates a feature vector as follows:

\[
F(L) = \begin{bmatrix}
f_{\text{linguistic}}(L) \\
f_{\text{emotional}}(L) \\
f_{\text{cognitive}}(L) \\
f_{\text{structural}}(L)
\end{bmatrix}
\]

Where:
\begin{align*}
f_{\text{linguistic}} & : \text{Linguistic attributes of } L \\
f_{\text{emotional}} & : \text{Emotional tone and sentiment in } L \\
f_{\text{cognitive}} & : \text{Cognitive complexity within } L \\
f_{\text{structural}} & : \text{Structural and grammatical aspects of } L
\end{align*}

These feature vectors facilitate nuanced comparisons between the 'control' and 'depression' groups, enriching our understanding of their linguistic and emotional preferences.

The mean of each LIWC feature for both classes was calculated, followed by the computation of the Mean Difference (MD):

\begin{equation}
\text{MD} =  \frac{1}{m} \sum_{i=1}^{m} x_{i, \text{Control}} - \frac{1}{n} \sum_{j=1}^{n} x_{j, \text{Depression}} 
\end{equation}

Here, \( m \) and \( n \) represent the sample sizes of the "control" and "depression" classes, respectively. \( x_{i, \text{Control}} \) and \( x_{j, \text{Depression}} \) are the specific feature values in each class.

Finally, we identified the top 10 features with the most significant mean differences, specifying the directionality to indicate whether the "control" or "depression" group is favored.

\begin{itemize}
    \item \textbf{Word Count (WC)} (+360.38): More complex narratives in the depression group.
    \item \textbf{Words Per Sentence (WPS)} (-14.73): Simpler sentences in the depression group.
    \item \textbf{Clout} (+2.70): Greater social influence in lyrics favored by the depression group.
    \item \textbf{Punctuation (AllPunc)} (-1.46): Less formal writing styles in the depression group.
    \item \textbf{Function Words (Function)} (+1.25): More complex cognitive processing in the depression group.
    \item \textbf{Linguistic Processes (Linguistic)} (+1.05): Intricate language use in the depression group.
    \item \textbf{Dictionary Words (Dic)} (+1.00): Broader vocabulary in the depression group.
    \item \textbf{Comma Usage (Comma)} (-0.84): Less intricate sentences in the depression group.
    \item \textbf{Big Words (BigWords)} (+0.75): Preference for complex words in the depression group.
    \item \textbf{Authenticity (Authentic)} (+0.71): Slight preference for authentic lyrics in the depression group.
\end{itemize}

%\subsubsection{Roberta}

\subsection{Network Analysis}
To answer \textbf{RQ4}, the question regarding network relationships, we analyze music sharing patterns, adopting a network-centric approach to elucidate the underlying structures and associations that may characterize the distinctions between depressed and non-depressed individuals. The core objective is to ascertain whether there exists a notable clustering propensity among songs shared by individuals with depression within distinct communities of music playlists. Our analysis is based on the Million Spotify Playlist Dataset \cite{SpotifyPlaylist}.

Playlists were pruned to retain only those songs that have a corresponding presence in our pre-established music-sharing dataset. A playlist is classified as a ‘depression playlist’ if a majority, i.e., over 50\%, of its comprising songs have been shared by individuals identified with depression on Twitter. In the constructed network, nodes epitomize individual playlists. Edges, on the other hand, are illustrative of the existence of shared songs between pairs of playlists, with the weight of each edge being indicative of the quantity of shared songs. We used community detection on this network to find potential patterns and associations inherent to the songs shared by individuals with depression. 

Community detection analysis yielded distinct patterns. It unveiled diverse degrees of dominance by depression and non-depression playlists, indicative of heterogeneous user behaviors and music preferences. For Community detection we used the Louvainalgorithm\cite{traag2019louvain}. 40 communities were detected in total. Of these, 9 communities have more than 10 nodes each. Among the 5 communities were dominated by the depressed population, while 2 communities were dominated by control. 2 communities showed almost equal amounts of depressed and control individuals. 

\begin{figure*}[htbp]
    \centering
    
    \begin{subfigure}[b]{0.2\textwidth}
        \includegraphics[width=\textwidth]{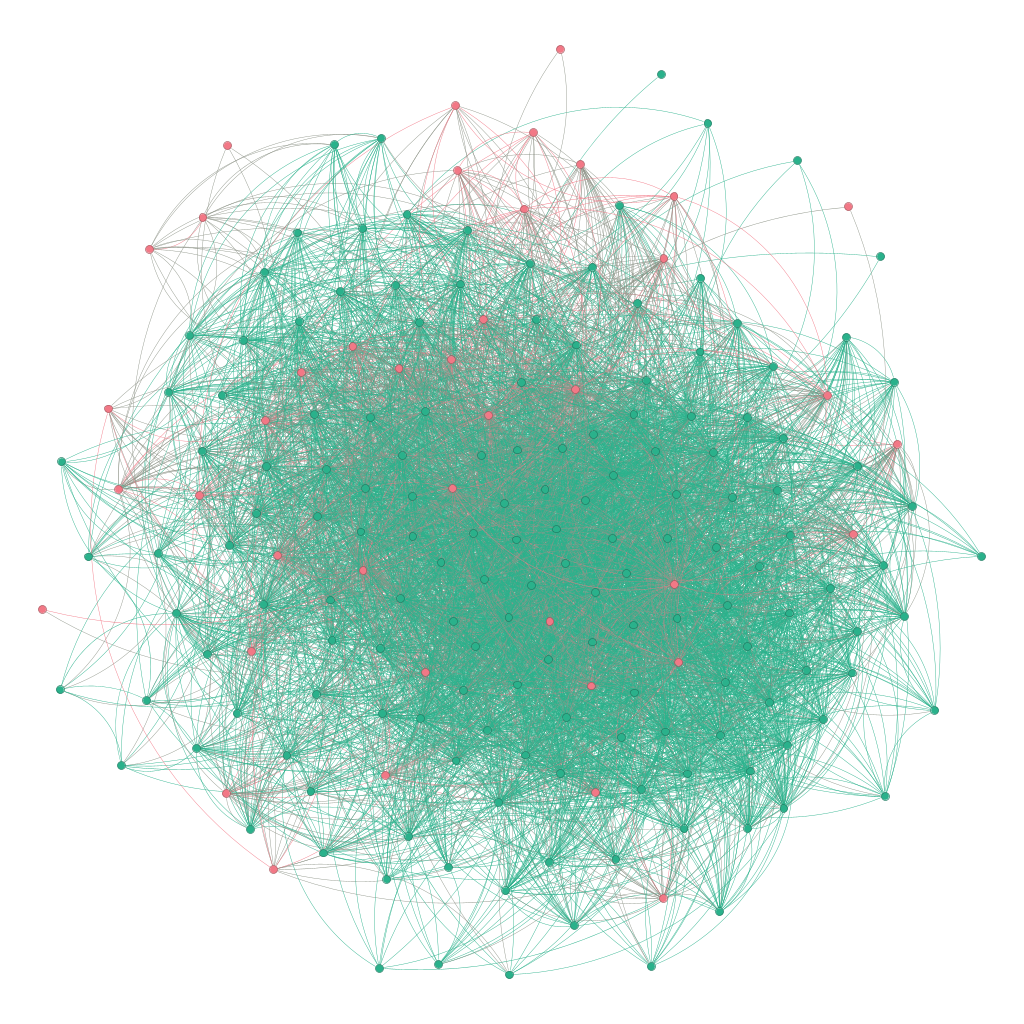}
        \caption{Community 1}
        \label{fig:community1}
    \end{subfigure}
    %\hfill % add some horizontal spacing between the two images
    \begin{subfigure}[b]{0.2\textwidth}
        \includegraphics[width=\textwidth]{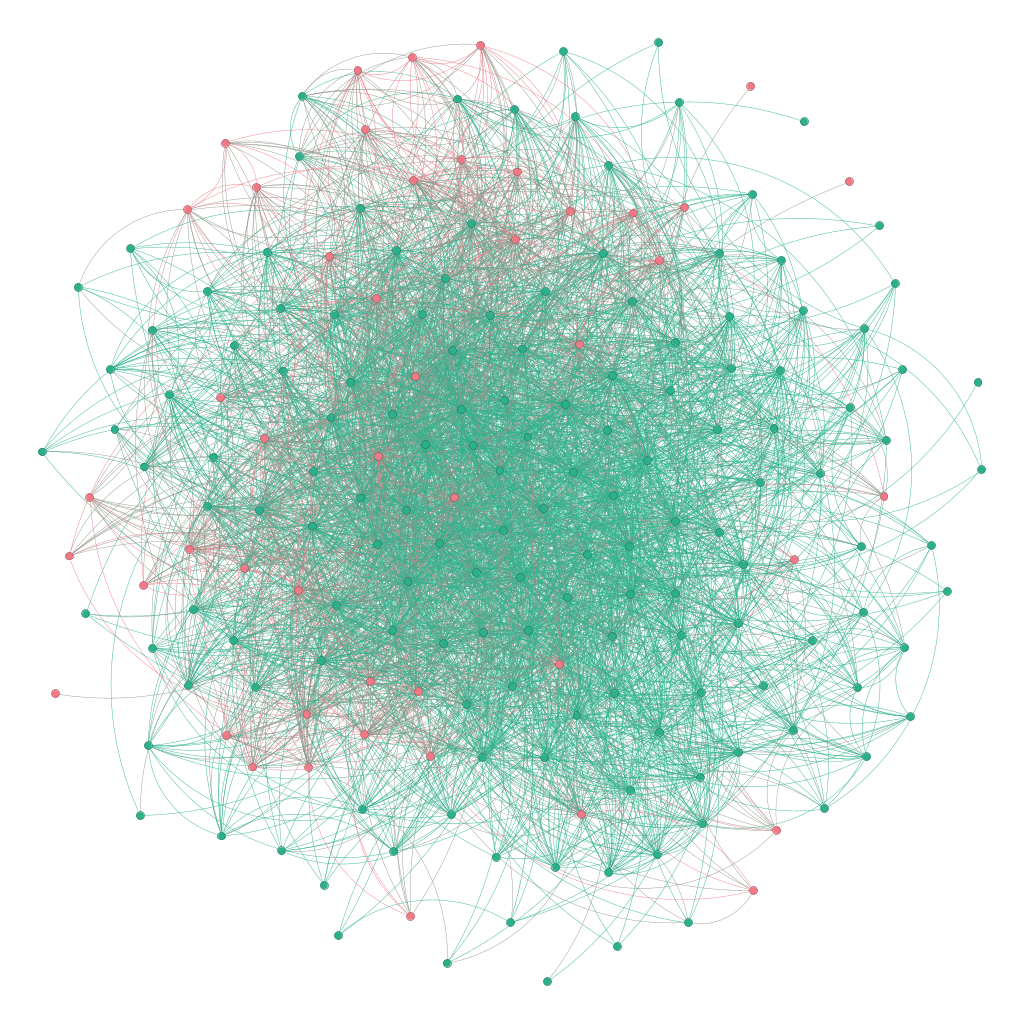}
        \caption{Community 4}
        \label{fig:community4}
    \end{subfigure}
    \begin{subfigure}[b]{0.2\textwidth}
        \includegraphics[width=\textwidth]{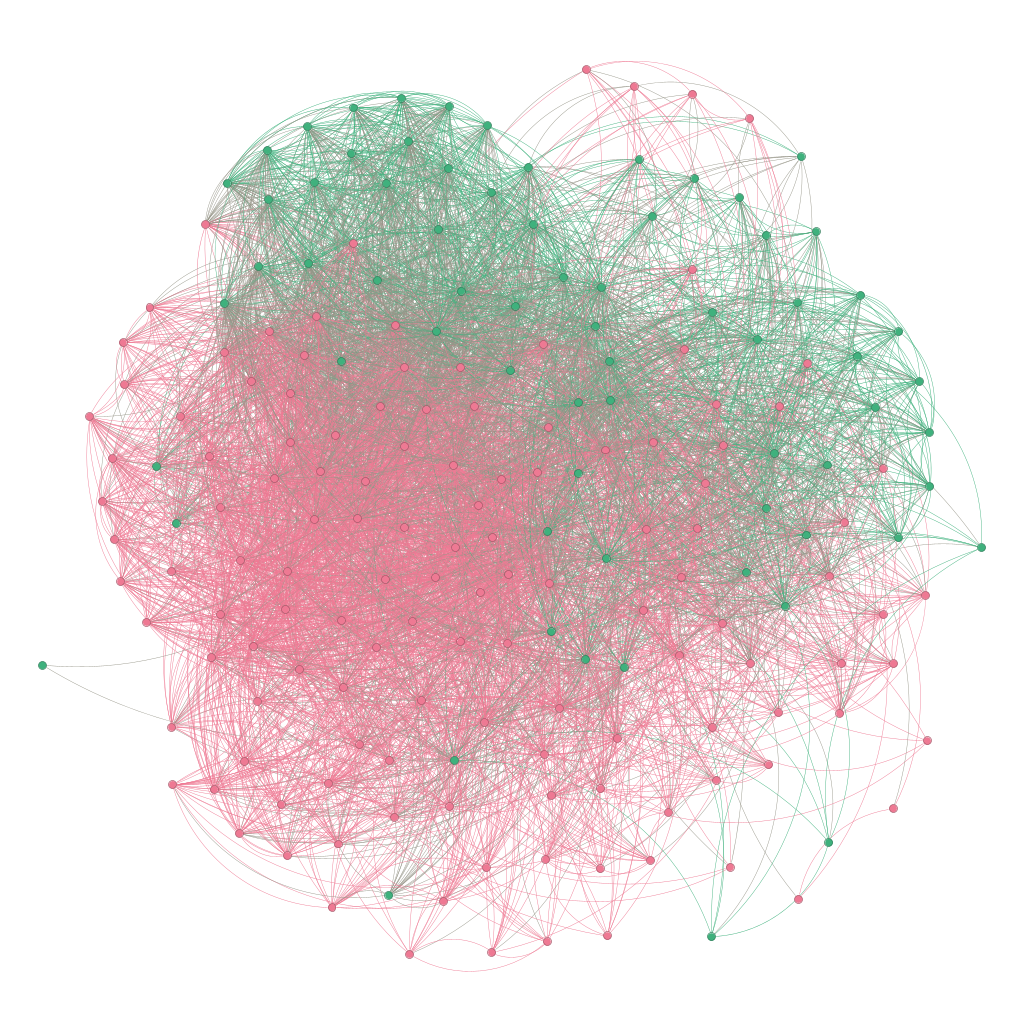}
        \caption{Community 5}
        \label{fig:community5}
    \end{subfigure}
    %\hfill
    \begin{subfigure}[b]{0.2\textwidth}
        \includegraphics[width=\textwidth]{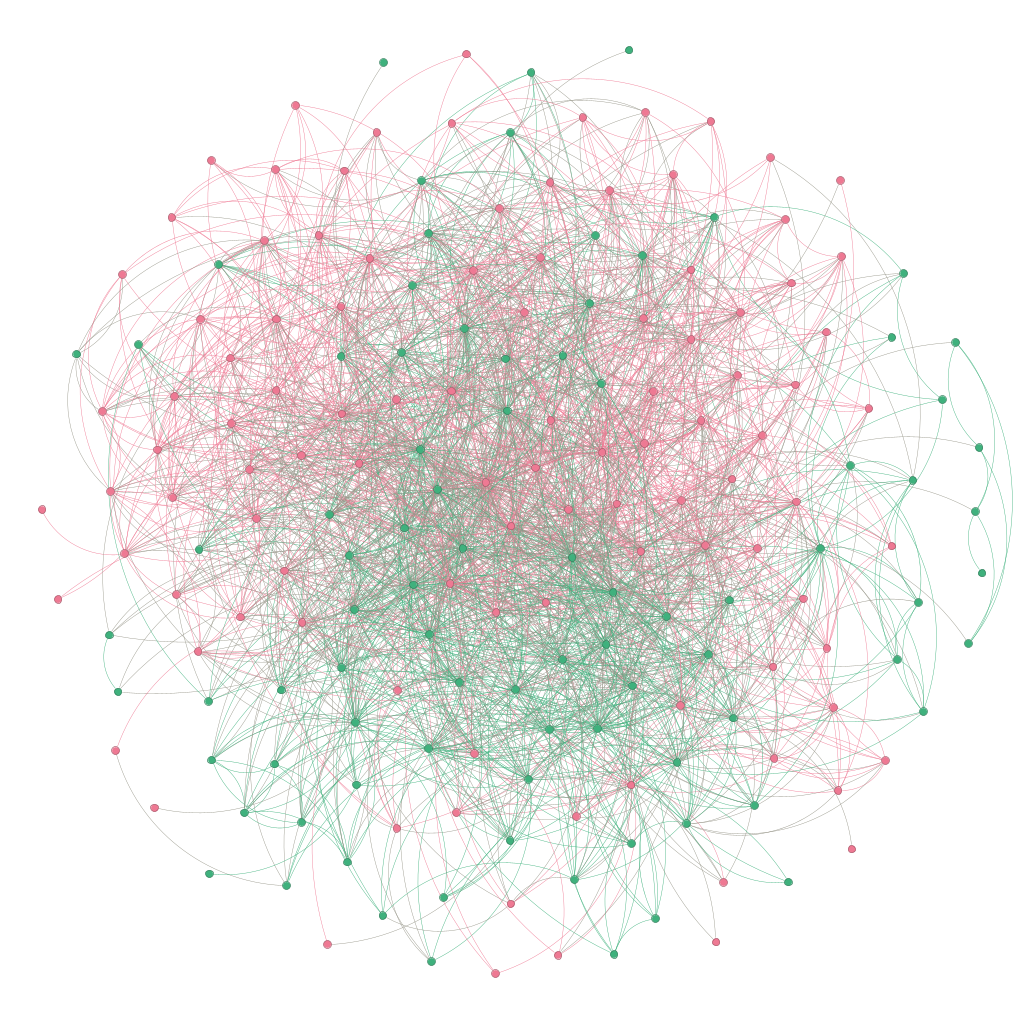}
        \caption{Community 0}
        \label{fig:community0}
    \end{subfigure}

    %\vspace{1em} % add some vertical spacing between the two rows
    
   % \begin{subfigure}[b]{0.3\textwidth}
      %  \includegraphics[width=\textwidth]{Sections/Community_Detection/community5.png}
      %  \caption{Community 5}
      %  \label{fig:community5}
    %\end{subfigure}
    %\hfill
    %\begin{subfigure}[b]{0.3\textwidth}
     %   \includegraphics[width=\textwidth]{Sections/Community_Detection/community0.png}
     %   \caption{Community 0}
     %   \label{fig:community0}
   % \end{subfigure}
    
  %  \caption{Communities detected in our network analysis.}
   % \label{fig:communities}
\end{figure*}

In the ensuing analysis, playlists associated with depression are represented in green, while those unrelated to depression are depicted in pink. Communities 1 and 4 are archetypal of clusters dominant in depression, as illustrated in Figures \ref{fig:community1} and \ref{fig:community4}. This observation is contrasted by Community 5 (Figure \ref{fig:community5}), which is characterized by a dominance of non-depression playlists. Community 0, highlighted in Figure \ref{fig:community0}, presents a balanced amalgamation of both depression and non-depression playlists.

The dichotomy in music sharing patterns, as evidenced through the network analysis, provides valuable insights and augments our comprehensive understanding of the behavioral and preferential disparities linked to depression. Each community, delineated by distinct characteristics, affirms the premise that music preferences and sharing behaviors are intrinsically associated with the mental and emotional states of individuals. This investigation not only amplifies the existing body of knowledge but also lays a substantive foundation for subsequent research endeavours aiming at the nexus of mental health, digital behaviors, and music preferences.

We additionally leveraged Latent Dirichlet Allocation (LDA) and community detection to identify the prevailing topic within each community. Intriguingly, communities 3 and 4, which are dominated by depression playlists, were overwhelmingly governed by the Emotional Reflection topic. Similarly, community 8, heavily associated with depression playlists, was primarily dominated by Religious or Biblical Language topic. These insights are consistent with preceding statistical analyses, indicating a higher prevalence of Emotional Reflection and Religious or Biblical Language topics within songs related to depression.

\section{Discussion}
\subsection{Summary of Findings}

In this work, we analyzed the difference of music sharing between depression and non-depression group from the perspectives of musical features, lyrics and musical networks. Statistical analysis of musical features showed that there are songs shared by depression are higher in energy, mode, acousticness, instrumentalness, and duration, and lower in speechiness, time signature, popularity and danceability. Lyric analysis showed that depression music sharing have higher proportion in music recording, religious or biblical language, everyday life, entertainment industry, recollection and creativity, everyday experiences, emotional reflection, and everyday conversations topics, yet lower proportion in explicit or RAP language, expressive language, musical artists and collaborations, and personal relationships topics. Additionally, individuals with depression may prefer songs with shorter and simpler sentences, more assertive or confident language. In addition, Network analysis revealed that depression playlists cluster together, suggesting a network connection between depression songs. Emotional reflection and religious or biblical language are dominant topics in depression playlist communities, which are different from non-depression playlist communities. This explorative analysis not only heightens our understanding of the nuanced relationships between music preferences and mental health but also illuminates pathways for cultivating sensitive, adaptive interventions and technologies in mental health discourse and practice.

\subsection{Implications}

For RQ1, our work showed that several musical patterns differ between depression and non-depression users. Our result is consistent with previous research on music and psychology, showing that depression patient tend to choose music with higher energy \cite{howlin2021patients}. Moreover, previous research indicated that higher energy decrease the perception of relaxing for songs \cite{baltazar2020songs}. However, as mental health research suggests that relaxation can significantly reduce the level of depression \cite{jia2020relaxation}, it raises an intriguing paradox: while individuals with depression are drawn to high-energy music, which might be less perceived as relaxing, they may unintentionally be selecting music that doesn’t optimize their psychological well-being, given the documented benefits of relaxation for mitigating depressive symptoms. Moreover, our results found that songs shared by depression individuals have lower popularity. Previous research on depression also showed that preference for subculture elements (such as anime, goths, punks) and less popular interests are positively associated with depressive symptoms \cite{liu2022does, bowes2015risk,bevsic2009punks}. This preference for less popular songs might stem from a variety of factors: the depressive mood might be reinforced by subcultures \cite{brown2017non}, and people with depression feel disconnected with popular culture \cite{liu2022does}. Additionally, reduced preference for speechiness is coherent with the genres and lyrics analysis showing that songs shared by depression users are less likely to be rap and contain explicit and rap languages. This illustrates a consistent connection between the analysis from different aspects in our work. 
%It is also worth noting that we did not detect the difference in valence between depression and non-depression groups. This is contradicted with previous research showing that individuals with depression are especially likely to listen to music that expresses negative valence because it matches their chronic mood state \cite{wilhelm2013blue}.

For RQ2, our work qualitatively showed that the artists and genres commonly shared by depression and non-depression individuals are different. The result shared common ground with previous work indicating that soul music listening(hip hop and R\&B) is a predictor of lower depression levels \cite{miranda2008personality}. Hip-hop and R\&B, often associated with expressions of resilience, empowerment, and an outlet for navigating through socio-emotional challenges, might not resonate with depression individuals' emotional and psychological state. 

%The communal and socially engaging attributes of hip-hop and R\&B culture, such as communal singing, dancing, and social commentary, might not coherent much with depressed individuals, potentially due to a diminished sense of belonging and connection, which are often impacted by the emotional and social withdrawal characteristic of depressive states \cite{cockshaw2014depression}. 

For RQ3, our results demonstrated that the topic and language use in depression and non-depression groups might be different. First, depression-prone individuals are more likely to share songs with topics in everyday life and emotional recollection. This is coherent with psychology research indicating that those with depression report more rumination than the control group \cite{olatunji2013specificity}. However, rumination can lead to negative consequences in mood: rumination exacerbates and prolongs depressed mood \cite{nolen1993effects}, and rumination makes it harder for patients to recover from psychological disorders \cite{watkins2020reflecting, morrison2008systematic}. We hypothesize that listening to songs with emotional recollection topic may induce ruminative thoughts, which, in turn, exacerbate depression. Moreover, the LIWC result indicated that depressed individuals prefer lyrics with short sentences. This is probably because depression decreases the cognitive load \cite{nikolin2021investigation}. Therefore, short sentences are easier to be perceived and liked by depressed individuals. 

Finally, for RQ4, we found that most of the communities display polarized with respect to either depression or non-depression playlists, which implies that the depression playlists may share similar music preference and cluster together. This may lead to the echo chamber effect, making depressed individuals being suggested by more depression songs, and further exacerbate and prolong depressed mood. Moreover, consistent with our topic modeling result, our network analysis also showed that the dominant topics of depression and non-depression communities are different. The dominance of religious and biblical languages in depression community is coherent with both LDA result and previous research showing that depressed individuals engaged more with expression of religious involvement on social media \cite{de2013social}. In summary, our work validate that the music preference is different between depression and non-depression group from both single song and playlist level. Utilizing both LDA and community detection serves to corroborate our findings and enhance the validity of the result. 

We discern crucial implications across two pivotal dimensions. Firstly, the characteristic features of individuals with depression in relation to their music preferences can be observed. On a social level, this group demonstrates a propensity for songs that may not be widely recognized or popular, possibly reflecting a diminished sense of belongingness and reduced proclivity for social engagement. This is also emphasized by their inclination away from soul music, often imbued with themes of resilience, empowerment, and socialization. From a cognitive perspective, a preference for shorter sentences may indicate a desire for easily understood content, potentially deriving from a lower cognitive load. Secondly, we delve into the consequences engendered by these specific musical preferences. Individuals with depression tend to gravitate towards high-energy music, paradoxically offering less relaxation, and exhibit an avoidance of genres articulating resilience. Moreover, a preference for songs with emotional reflection topics may inadvertently foster rumination, further entrenching depressive states. This is not only evident in the lyric analysis in the network analysis. Incorporating all aforementioned observations, the musical preferences of depression individuals inadvertently forge a self-reinforcing echo chamber, where the music perpetuates and potentially exacerbates prevailing depressive moods and emotions. This echo chamber effect is manifested not merely through individual listening choices but is further perpetuated by the recommendation algorithms that may suggest additional depression playlists to listeners. Thus, recommendation algorithms, influenced by these patterns, may unwittingly suggest additional depression-congruent content to listeners, thereby sustaining and potentially intensifying the resonance with depressive moods and emotions through continued exposure to thematically similar music. 

Based on the result, we also suggest that music application develop algorithms or frameworks to gauge the emotional state of listeners, by analyzing their music listening patterns over time, thus providing an unintrusive way to monitor emotional well-being. Music application can inquire the mental health condition of their users and recommend songs or playlists with less depressive features, and explore the efficacy of such tailored musical interventions and whether guiding listening habits can serve as a viable strategy in emotional self-regulation or mental health management. In addition, the answer to the question "How can recommendation algorithms avoid negative reinforcement and echo chamber effect?", needs to be explored. These ideas may help developing holistic mental health supporting systems.

%As for RQ4, we found that most of the communities are polarized by depression or non-depression playlists, which imply that the depression playlists may share similar music preference and cluster together. This may lead to the echo chamber effect, making depression people being suggested by more depression songs, and further exacerbate and prolong depressed mood. Moreover, consistent with our topic modeling result, our network analysis also showed that the dominate topic of depression and non-depression communities are different. This also validate that the music preference is different between depression and non-depression group from both single song and playlist level. 

\subsection{Limitation and Future Work}
We acknowledge our study has several limitations, some of which point to intriguing future research directions. First, this study focused on the overall profile and preferences of depressed and non depressed users, while a potential future work can be eliminating the overlapping features and analyzing the distinctive unique features for each of the users classes. Second, since we do not have the playlist information from depression/non-depression users, we cannot analyze the actual music preference connection between depressed individuals in the network analysis. Instead, we have to analyze in an indirect way that regards playlists with a majority of songs shared by depressed individuals as depression playlists. A potential future direction could focus on collecting social network data with depression playlists to further study this question.

%Even though we have playlist, we cannot generate the actual connection between them. proxy, indirect way
%

\bibliographystyle{ACM-Reference-Format}
\bibliography{main}

\appendix

\end{document}